\documentclass[preprint,authoryear,5p]{elsarticle}
\usepackage{amssymb}
\usepackage{amsmath}
\usepackage{amsfonts}
\usepackage{url}
\usepackage{hyperref}
\usepackage{graphicx}

\journal{NeuroImage}

\begin{document}

\begin{frontmatter}

\title{Low-Rank + Sparse Decomposition (LR+SD) for EEG Artifact Removal}

\author[uclamath]{J.~Gilles}
\ead{jegilles@math.ucla.edu}

\author[uclamath]{T.~Meyer}
\ead{euphopiab@gmail.com}

\author[uclaneuro]{P.K.~Douglas}
\ead{pamelita@ucla.edu}

\address[uclamath]{Department of Mathematics, University of California, Los Angeles, 520 Portola Plaza, Los Angeles, CA 90095, USA}
\address[uclaneuro]{Jane \& Terry Semel Institute for Neuroscience \& Human Behavior, David Geffen School of Medicine, University of California, Los Angeles, California, USA}

\begin{abstract}

Concurrent EEG-fMRI recordings are advantageous over serial recordings, as they offer the ability to explore the relationship between both signals without the compounded effects of 
nonstationarity in the brain.  Nonetheless, analysis of simultaneous recordings is challenging given that a number of noise sources are introduced into the EEG signal even after MR 
gradient artifact removal with balistocardiogram artifact being highly prominent.  Here, we present an algorithm for automatically removing residual noise sources from the EEG signal in a 
single process using low rank + sparse decomposition (LR+SD). We apply this method to both experimental and simulated EEG data, where in the latter case the true EEG signature is known. 
The experimental data consisted of EEG data collected concurrently with fMRI (EEG-fMRI) as well as alone outside the scanning environment while subjects viewed Gabor flashes, a perceptual 
task known to produce event related power diminutions in the alpha spectral band.  On the simulated data, the LR+SD method was able to recover the pure EEG signal and separate it from 
artifact with up to three EEG sources. On the experimental data, LR+SD was able to recover the diminution in alpha spectral power that follows light flashes in concurrent EEG-fMRI data, 
which was not detectable prior to artifact removal. At the group level, we found that the signal-to-noise ratio was increased ~34\% following LR+SD cleaning, as compared independent 
component analysis (ICA) in concurrently collected EEG-fMRI data. We anticipate that this method will be highly useful for analyzing 
simultaneously collected EEG-fMRI data, and downstream for exploring the coupling between these two modalities.

\end{abstract}


\end{frontmatter}

\section{Introduction}
\sloppy

Simultaneous EEG-fMRI recordings are increasingly used to study the rich temporal dynamics of the human brain noninvasively.  Each modality offers complementary benefits in signal detection: 
EEG scalp recordings provide high density temporal data yet resolves over diffuse spatial regions in the source domain, while fMRI signals provide high spatial precision, and yet evolve 
slowly due to temporally smoothed hemodynamics \citep{Daunizeau2009}. The idea of creating a principled method for EEG-fMRI fusion is highly attractive \citep{ValdesSosa2009}, particularly 
given its promise in localization of epileptic foci \citep{Thornton2011,Grouiller2011}. However, analysis of these multimodal data have proven quite challenging, given a number of artifacts 
introduced - particularly into the EEG signal - during concurrent recordings.

EEG recordings putatively reflect the superposition of electric dipoles associated with synchronous activity from neural populations measured at the scalp \citep{Buzsaki2012,Dahne2012}. 
When collected outside the MR scanning environment, these signals are corrupted by a series of artifacts including: line noise (50\slash 60 Hz electrical noise), electromyogram (EMG) 
activity, and electro-oculogram (EOG) noise. In the case of simultaneous EEG-fMRI experiments, the artifacts are even more complicated.  The most prominent signal contamination arises due to 
the switching of magnetic fields. However, a number of averaging based template-subtraction methods exist which are largely effective at removing this periodic gradient artifact 
\citep{Goldman2002,Ritter2006}.

Although smaller in amplitude, a number of additional quasi-periodic artifacts cause electromagnetic interference in the MR environment that are not easily removed using template based 
methods. The ballistocardiogram (BCG), generated as EEG electrodes move due primarily to pulsatile head-rotation motion during the cardiac cycle \citep{Mullinger2013}, is particularly 
problematic \citep{Srivastava2005}, and presents broadly in the spectral frequencies of interest in EEG (0.5-25 Hz) \citep{Debener2008}. Similarly, the magnetic cryo-pump can induce noise 
with high trial-by-trial variability, and is therefore turned off typically during concurrent EEG-fMRI acquisition \citep{Mullinger2013}. The presence of these artifacts can dramatically 
change the spectral properties of the signal, and obscure ability to perform trial-by-trial analyses.

\subsection{EEG Artifact Removal Techniques}

Many artifact removal techniques have been developed in the last decade to address these issues. For instance, in \cite{Allen1998}, the authors propose a template-based BCG artifact removal 
method called AAS (Average Artifact Subtraction). First, they detect timings when R peaks occur in the ECG (electrocardiogram) signal and extract several ``small'' portions of the EEG signal 
centered to these timings which are averaged to create a template for subtraction of the BCG artifact. In \cite{Allen2000}, the authors combine the AAS method and an Adaptive Noise 
Cancellation (ANC) procedure synchronized with the slice-timing signal in order to remove residual artifacts. Independent component analysis (ICA) based blind source separation techniques 
have also been used for BCG removal \citep{Makeig1996,Jung1998,Jung2000}. Effectively, ICA decomposition of the $N$ EEG channels is performed and the corrected EEGs are obtained by 
recomposing the ICA components without the ``artifact components,” and the process of how a component is designated as artifact or not can be manual or automatic. In \cite{Mantini2007}, the 
authors propose a method which combines AAS and ICA techniques to clean the EEG signals. A temporal principle component analysis (PCA) method has been used to create a series of artifact 
templates, termed optimal basis set (OBS) \citep{Niazy2005}, and has also been used in combination with ICA \citep{Debener2007}. In \cite{Zou2012}, clustering is applied to IC components 
to separate the artifact components from the useful ones. In \cite{Sun2009}, the authors adapt a blind source separation algorithm called Maximum Noise Fraction (MNF), originally developed 
for multispectral imaging, to remove the BCG artifact. 

Another category of techniques is based on adaptive filtering methods. In \cite{Masterton2007}, the authors record the BCG and body movements separately. Then they use an adaptive filtering 
technique called ``Multi-Channel Recursive Least Square,” a particular case of Kalman filtering, to adaptively remove the recorded movements from the EEG signals. This approach seems to 
outperform AAS techniques but requires extra movement recordings. In \cite{Correa2007}, the authors simulate the line noise and use extra recording of ECG and EOG as input of three adaptive 
filters. The first one remove the line noise, the second one the ECG artifact and the last one the EOG artifact. Each filter is a finite impulse response filter where their coefficients are 
continuously updated by a Least Mean Square algorithm. Nonetheless, most artifact removal methods generally require either a priori knowledge in the form of user identification of artifact 
components or an artifact template, which may be time consuming or unavailable. 

Here, we introduce a new algorithm for removing artifact from EEG signals that uses low rank + sparse decomposition (LR+SD). To do so, we propose a mathematical model based on a reasonable 
experimental assumption that artifact components will be mathematically expressed differently than the data themselves. Importantly, this method obviates the need for any reference or 
template artifact signal.  As such, the combined effects of many types of artifacts can be removed in a single decomposition without the need for manual identification of artifact components 
in the data.  We assess the utility of this new algorithm first using simulated EEG data with a known solution.  Next, we apply this method to real experimental EEG data collected 
concurrently in the scanner room with fMRI data while subjects performed a visual perception task. We then compare our novel LR+SD method to conventional ICA based artifact removal 
techniques to EEG data collected serially outside of the scanner on this same visual task.

\section{Methods}

\subsection{Overview}
Our method in brief is as follows.  We first describe our simulated EEG data set with a known solution. We then describe our stimulus paradigm, our method for concurrent EEG-fMRI data 
collection, and our serially collected EEG data set collected outside of the MR scanning environment using the same stimulus paradigm. We used both in and out of scanner data for testing our 
novel artifact removal technique experimentally. We then describe our LR+SD method in detail, and methods used for comparing the performance of this algorithm to existing methods.

\subsection{Simulated Dataset}

In general, there is no ground truth EEG signal when data are empirical, making it difficult to assess the utility of artifact-removal algorithms. We therefore created a simulated dataset by 
using the free BESA (Brain Electrical Source Analysis) Simulator \footnote{Available at \url{http://www.besa.de/updates/besa_simulator/}} to generate ``pure'' EEG 
signals corrupted by known artifacts.
We prepared two simulated datasets of varying complexity using forward modeling with a spherical head model.  Data were simulated using the``Adult cr80'' EEG sensor model, sampled at 125Hz 
over 10000 points. In dataset I, we simulated data for a single dipole source near the cortical surface in the visual region. In dataset II, three dipoles sources located in distributed brain 
areas were used to simulate data. 
A Hanning filter was applied to all simulated sources. For data set I, we simulated data for a 20~nAm sine wave at 10Hz occurring in the interval $[15-17]$s, located at 
$(-0.219,-0.646,0.036)$ with orientation 
$(0.479,-0.336,-0.811)$ (visual cortex). In simulation II, the three sources included: one 4~nAm sine wave at 12Hz in the interval $[5-10]$s at position $(-0.219,-0.646,0.036)$ with 
orientation 
$(0.479,-0.336,-0.811)$ (visual cortex), one 3~nAm sine wave at 6Hz in the interval $[23-33]$s at position $(0.284,0.644,0.168)$ with orientation $(-1,0,0)$ (frontal lobe) and 
a 5~nAm sine wave at 20Hz in the interval $[53-56]$s at position $(-0.438,0.05,0.5)$ with orientation $(1,0,0)$ (primary motor cortex).
In order to add realistic noise to the data, we used ECG, EMG, and right and left EOG reference artifact recordings extracted from the free sample of the SHHS Polysomnography 
Database\footnote{Available at \url{http://physionet.org/physiobank/database/shhpsgdb/}} (see \cite{Physionet,SHHS1,SHHS2,SHHS3} for more information). 
These reference artifacts were normalized and added to the pure simulated EEGs using randomized mixing coefficients accordingly to a uniform distribution.

\subsection{Experimental Data: Participants and Stimuli Design}
A total of ten healthy individuals (ages 23-30, twelve male) without history of neurological disease were recruited to participate in this study, which was approved by the UCLA Institutional 
Review Board. Each individual provided written informed consent at the Staglin Center for Cognitive Neuroscience prior to participation in the study. The experiment consisted of concurrent 
EEG-fMRI data collection, followed by single modality EEG data collection outside the scanner suite. 

Concurrent data was collected in a dimly lit MR scanner room while a subject passively viewed visual stimuli, presented via an MR projector screen.  Stimuli consisted of Gabor element 
‘flashes’ presented radially against a black background. A total of ~140 flashes were presented to each subject for 500 msec each. We jittered the timing of the stimulus presentation with a 
minimum inter stimulus interval of 9 sec (mean 12.85 +/ -2.8 sec).  We used a long ISI to allow plenty of time for the EEG alpha power to return to baseline following the flash stimulus. 
During the inter stimulus interval, subjects were asked to fixate on a low contrast small grey crosshair centered on the screen.  We selected this task, as it has been shown to generate 
reproducible activations in functionally defined brain regions across both modalities.  Light stimuli consistently elicit event related spectral perturbations (ERSPs) in EEG occipital 
channels \citep{Huettel2004}. 

\subsection{Data Acquisition: Concurrent EEG-fMRI and Single Modality EEG}
Simultaneous EEG/ fMRI data were recorded using a high density 256-channel GES 300 Geodesic Sensor Net (Electrical Geodesics, Inc.) with an EEG sampling rate of 500 Hz, and TR of 1sec. MRI 
clock signals were synced with EEG data collection for subsequent MR artifact removal. Functional scans were acquired using 3-T Siemens Trio MRI Scanner using echo planar imaging 
gradient-echo sequence with echo time (TE) of 25msec, repetition time of 1s, 6mm slices, 2mm gap, flip angle 90 degrees, with 3mm in-plane resolution, ascending acquisition. A high resolution 
structural image was collected using the magnetization-prepared rapid-acquisition gradient echo (MPRAGE) protocol (slice thickness 1.5mm, 1mm in plane resolution, flip angle 20 degrees, TE 
of 4.76, along with a T2 matched bandwidth scan to facilitate registration. 

For comparison purposes, we also collected EEG data using this same visual task outside the scanner environment. These data were collected in a dimly lit copper shielded room using an EGI 
256-channel Geodesic Sensor net with a 500 Hz sampling rate. 

\subsection{EEG Data Processing}

Initial EEG data preprocessing and artifact detection was performed using tools available in EGI NetStation.  EEG data first underwent MR artifact removal by subtracting an exponentially 
weighted moving average template, according to methods described in \cite{Goldman2002}. Parameters for cleaning were chosen based on matching the frequency power spectrum with that of 
data collected out of scanner. 

In order to compare our LR+SD method to established method, we performed either ICA or LR+SD on gradient-cleaned data.  ICA based BCG removal was performed by applying the InfoMax ICA 
algorithm using Brain Analyzer v.2.0.2 software (Brain Products). Noise artifacts were then identified manually by visually inspecting the time series of the ICs as well as the power 
spectrum across the scalp channels, as was done in \cite{Mantini2007}. Following either ICA or LR+SD cleaning, stimulus events were segmented 2 sec before and 10 sec after presentation of 
each stimulus. Subsequent artifact detection was accomplished using a gradient threshold of 400 $\mu$V. If $>$20\% of channels were marked as error, then that segment was excluded from further 
analysis, and if $>$30\% of segments were marked as bad, then that particular subjects data was omitted for that task. Lastly, data were DC detrended across the duration of each segment.

\section{Low Rank + Sparse Decomposition Method}
We denote $\{\tilde{f}_i(k)\}_{i=1}^N$ the set of recorded EEG signals. The index $i$ corresponds to the channel index, assuming we have 
a total of $N$ electrodes distributed over the scalp. Moreover, we assume that each signal is recorded over $K$ samples, i.e. $k\in\{1,\ldots,K\}$.
We consider $J$ (unknown) artifacts, and will 
denote them $\{f_j^A(k)\}_{j=1}^J$ where the index $j$ identifies different artifacts. The goal of the artifact removal procedure is then to retrieve cleaned EEG signals, 
$\{f_i(k)\}_{i=1}^N$.\\

In the proposed method we assume the following model: each recorded EEG channel is a linear combination of its cleaned version and the different artifacts. This model is equivalent to write
\begin{equation}
\tilde{f}_i(k)=f_i(k)+\sum_{j=1}^Ja_{ij}f_j^A(k),
\end{equation}
where the mixing coefficients $a_{ij}$ are unknown. 
In the following, we use a matrix formalism to model the global processes. To do so, we cast each EEG and artifact channels as columns of 
$K\times N$ matrices:
\begin{align}
\underbrace{\left(\begin{array}{cccc}
      | & | & & | \\
      \tilde{f}_1 & \tilde{f}_2 & \ldots & \tilde{f}_N \\
      | & | & & |
      \end{array}
\right)}_{\tilde{F}}&=\underbrace{\left(\begin{array}{cccc}
      | & | & & | \\
      f_1 & f_2 & \ldots & f_N \\
      | & | & & |
      \end{array}
\right)}_{F}\\ \notag
&+\underbrace{\left(\begin{array}{cccc}
      | & | & & | \\
      \tilde{f}_1^A & \tilde{f}_2^A & \ldots & \tilde{f}_N^A\\
      | & | & & |
      \end{array}
\right)}_{\tilde{F}^A},
\end{align} 
where $\tilde{f}_i^A(k)=\sum_{j=1}^Ja_{ij}f_j^A(k)$. The matrix $\tilde{F}$ contains all recorded EEG channels, $F$ the wanted cleaned EEG signals and $\tilde{F}^A$ contains the mixing of all 
artifacts. The latter can be written as $\tilde{F}^A=\sum_{j=1}^JF_j^A$ where
\begin{equation}
F_j^A=\left(\begin{array}{cccc}
      | & | & & | \\
      a_{1j}\tilde{f}_j^A & a_{2j}\tilde{f}_j^A & \ldots & a_{Nj}\tilde{f}_j^A\\
      | & | & & |
      \end{array}
\right).
\end{equation}
The key to our method is to notice that each matrix $F_j^A$ has its columns proportional to the same vector $\tilde{f}_j^A$ implying that $rank(F_j^A)=1$ and consequently 
$rank(\tilde{F}^A)\leq J$. 
Otherwise, the matrix $F$ should contain events resulting from “true” EEG data. In our case, we focus on event related spectral perturbations (ERSPs), which occur at specific times and affect 
a limited number of electrodes. Therefore, it is reasonable to assume that $F$ is a sparse matrix. Thus the artifact removal problem is equivalent to performing a low-rank + sparse 
decomposition of $\tilde{F}$.  The resulting sparse component therefore corresponds to the cleaned EEG signals. Such decomposition can be done by solving the following minimization problem:

\begin{gather}
(F,F^A)=\arg\min \|F^A\|_*+\lambda\|F\|_1 \\ \notag
\text{such that}\; \tilde{F}=F+F^A,
\end{gather} 
where $\|.\|_*$ denotes the nuclear norm of a matrix (i.e. the sum of its singular values), $\|.\|_1$ denotes the sum of the absolute value of the matrix entries and $\lambda$ is a positive 
parameter allowing us to control the rank of $F^A$. Such model, also called Robust PCA, was actively studied in the mathematics community, see for example \cite{Cai2008,Candes2009,Lin2009}. 
In our 
experiments we adopt the algorithm presented in\footnote{Matlab implementation available at \url{http://perception.csl.illinois.edu/matrix-rank/Files/exact_alm_rpca.zip}} \cite{Lin2009}. For 
artifact removal of the empirical data, the regularization parameter was selected by running a sweep across all possible ranks for a test subject. The selected regularization parameter was 
then applied to the remaining subjects.   

Upon publication, all code developed for this project will be available on the NITRC repository\footnote{At \url{http://www.nitrc.org}}. 

\section{Results}
The Time-Frequency representation (TFR) is widely used to visualize how the energy in different spectral bands is evolving with respect to some stimuli. In this section, we use the TFR, 
computed via a continuous wavelet transform (CWT) using the Morlet wavelet, to assess the efficiency of the proposed method.

\subsection{Results on Simulated EEG}
\begin{figure}[!ht]
\includegraphics[width=\columnwidth,height=0.7\columnwidth]{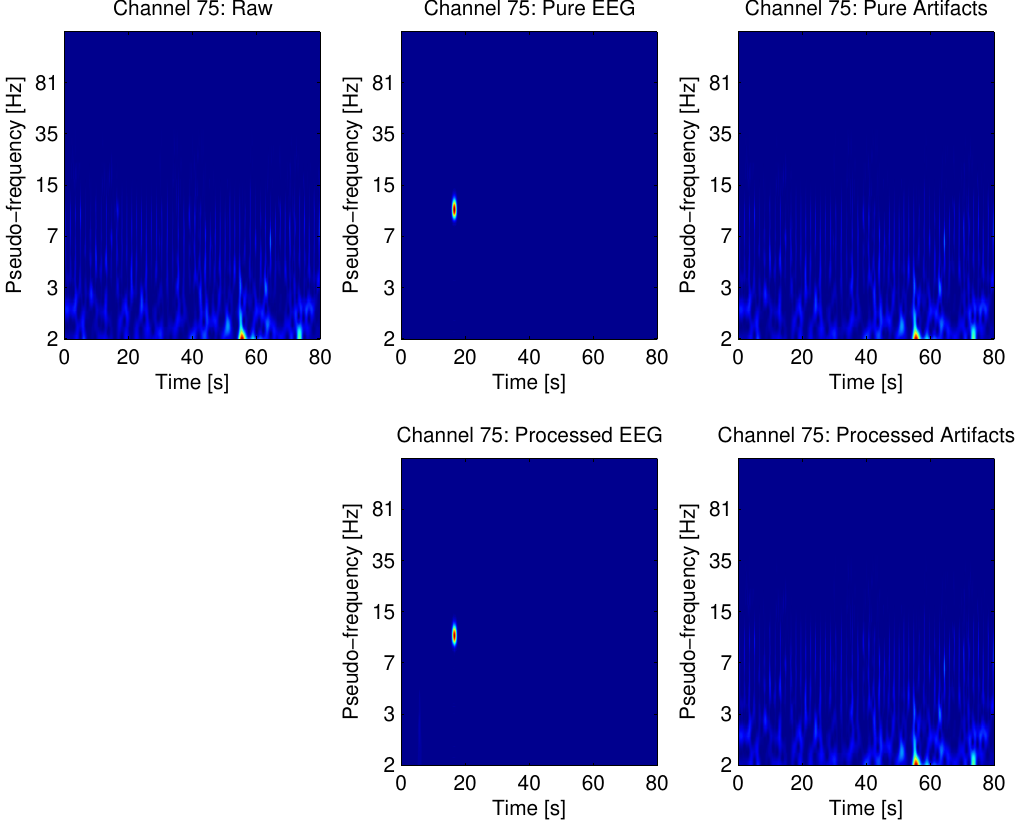}
\caption{Simulated Data Set I. A single dipole source located close to an occipital electrode channel is mixed with artifacts. (Top row) TFR of the original simulated raw data corrupted by 
artifact (left), the true EEG data alone (middle), and artifact alone (right). (Bottom Panel) TFR of LR+SD results, which separates the cleaned EEG (left) from artifact (right). Values are 
normalized to maximum for each display.}
\label{fig:single}
\end{figure}

In this section we illustrate the results obtained from applying low-rank + sparse decomposition (LR+SD) to the simulated datasets. In this case, we know the ``true'' EEGs and artifacts so 
our goal is to check if the proposed 
method is capable of retrieving both TFRs from the raw signals.
Figure~\ref{fig:single} shows the TFRs for the case of a single source. The TFRs corresponding to the pure and artifact signals are depicted in the two upper right plots while the TFR 
obtained from the raw EEGs (pure EEGs mixed with pure artifacts) is given on the upper left plot. 
Notice that the time-frequency 
energy corresponding to pure EEGs is nearly undetectable due to the artifact energy. The second row shows TFRs obtained from LR+SD.

\begin{figure}[!ht]
\includegraphics[width=\columnwidth,height=0.7\columnwidth]{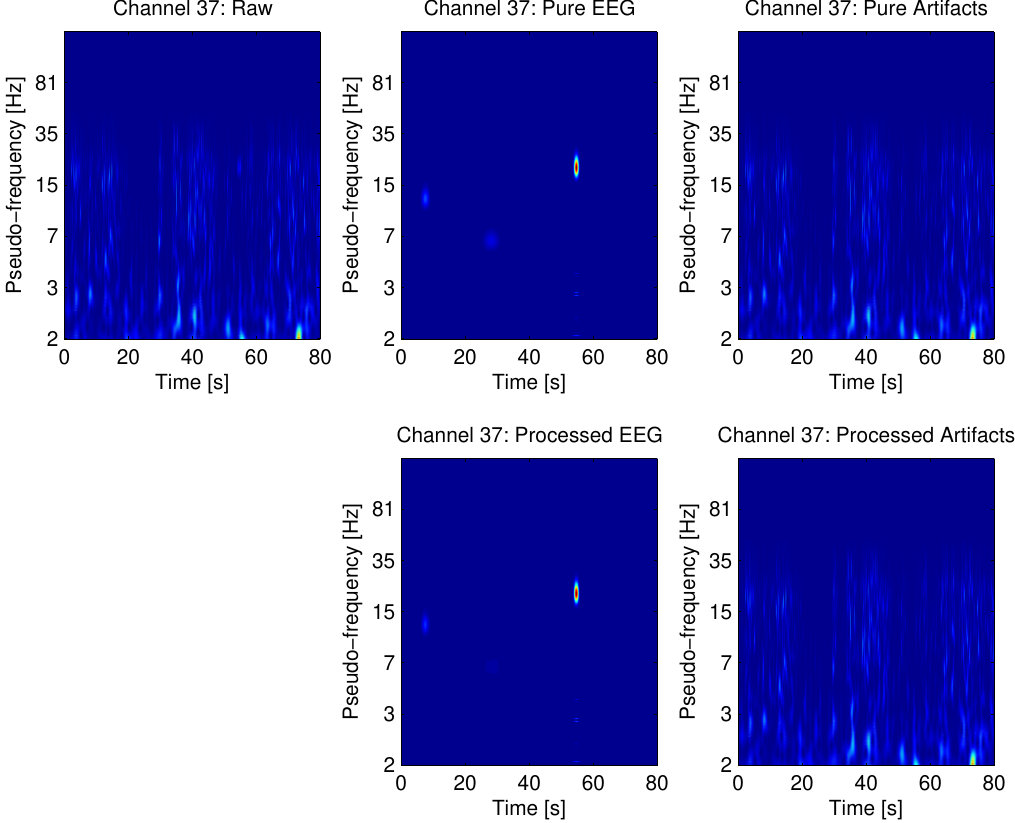}
\caption{Simulated Data Set II. Three sources measured with an electrode located close to the primary motor cortex (source 3). (Top Panel) CWT of original simulated data consisting of three 
dipole sources corrupted by ECG, EMG and right and left EOG template artifacts (left), the true EEG data alone (middle), and artifact alone (right). (Bottom Panel) LR+SD result TFRs 
separating the cleaned EEG (left) from artifact (right). Values are normalized to maximum for each display.}
\label{fig:three37}
\end{figure}

The second dataset contains multiple events at different times and in different area of the brain. Figure \ref{fig:three37} illustrates the 
obtained results (with 
the same plot layout as described for the single source case) from an electrode close to the activation areas. The signatures of each event are not visible in the raw EEGs' TFRs yet they are  
clearly visible in the sparse component.

In both the single and multiple source experiments the regularization parameter $\lambda$ had value $5.10^{-3}$ which resulted in ranks of 4 and 5 for the low-rank artifact components, 
respectively. The proposed method shows excellent results in separating the artifact parts from the EEG signals of interest.

\subsection{Results on Concurrent EEG-fMRI Acquisitions}

\begin{figure}[!h]
\includegraphics[width=\columnwidth]{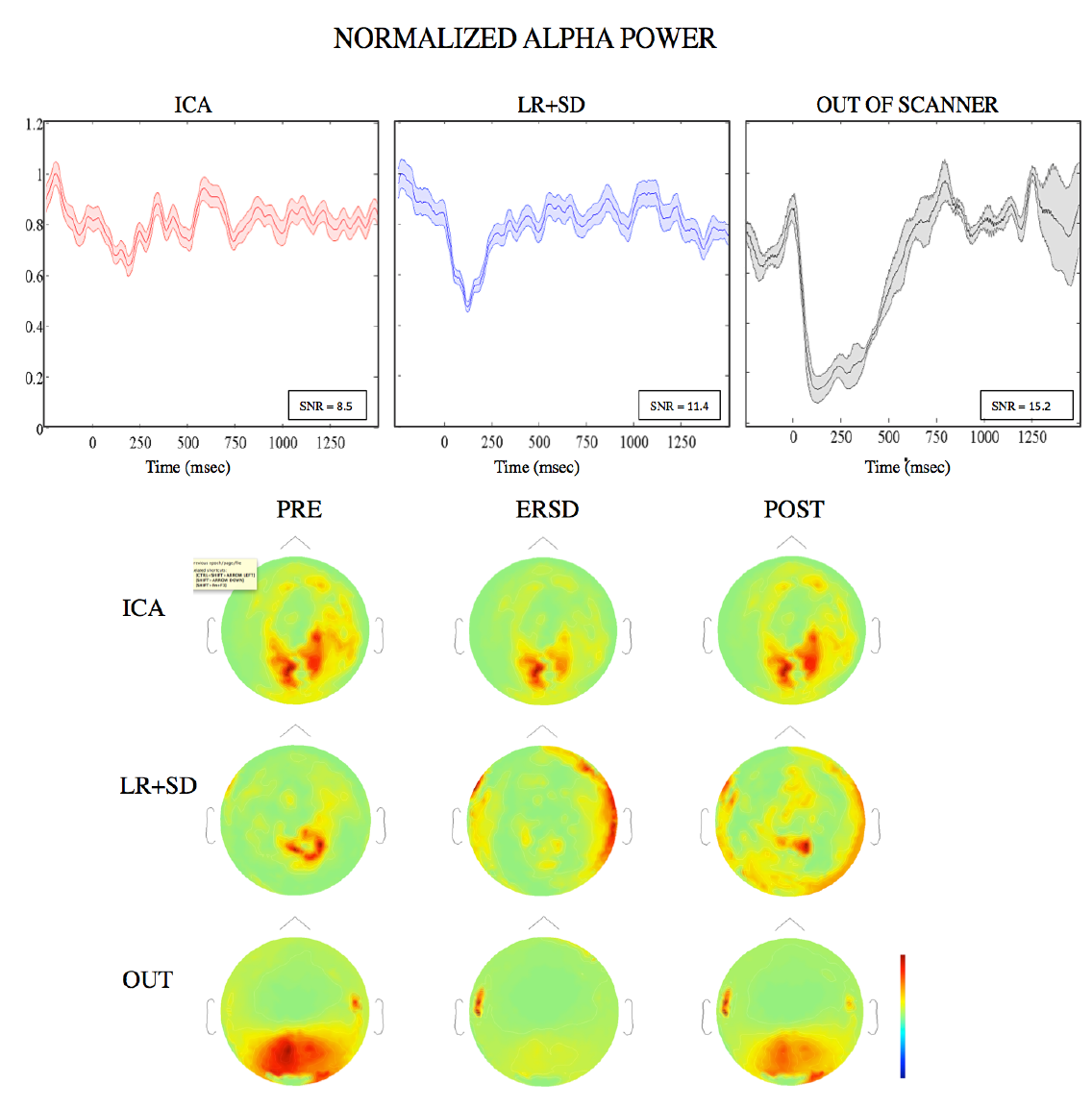}
\caption{Group Level Results comparing independent component analysis (ICA) and LR+SD based cleaning to EEG data collected outside of the MR scanner environment. (TOP PANEL) Normalized Alpha 
Power time courses derived from the ocular EEG channel averaged across all subjects plotted with mean +/- SEM for ICA, LR+SD and Outside Scanner Results. Signal-to-Noise rations are shown in 
the lower left corner for each result with the stimulus occurring at time equal to zero. (LOWER PANEL) Group level alpha power results projected topographically for 500msec prior to stimulus 
onset (PRE), 50 msec following stimulus onset(ERSD), and 500msec following stimulus onset (POST).}
\label{fig:pamfig}
\end{figure}

Group level ERSP results for LR+SD artifact removal of our experimental data are summarized and compared to ICA artifact removal as well as out of scanner data in Figure \ref{fig:pamfig}(a-d). 
Signal-to-noise ratio (SNR) was computed by calculating the ratio of the maximum absolute signal diminution in alpha power from 0 to 500 msec following stimulus presentation to the standard 
deviation of alpha power from the following 1000msec post stimulus. SNR was 8.5, 11.4, and 15.2 for ICA, LR+SD, and out of scanner data respectively. Figure \ref{fig:pamfig}(d) shows group 
level alpha spectral EEG data projected topographically for pre-stimulus (-250 msec), ERSP (50msec), and post stimulus (500 msec), with timings with respect to the stimulus occurring at time 
equal to zero. 

\begin{figure}[!h]
\includegraphics[width=\columnwidth]{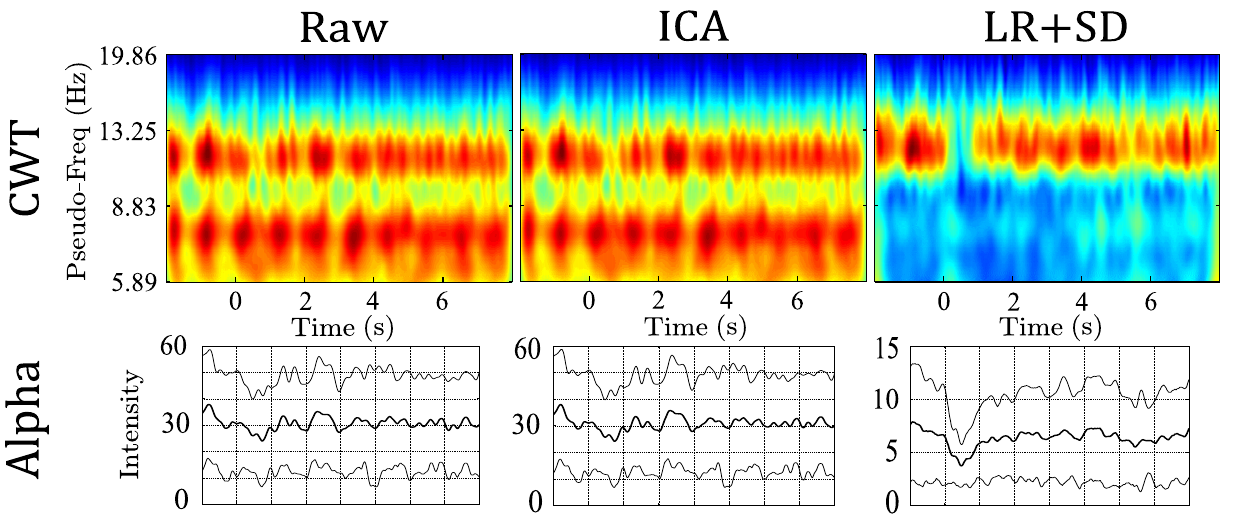}
\caption{Single-subject alpha band TFR averaged over stimuli. Columns denote (left) the raw MRI signal, (center) the ICA-cleaned signal (right) the sparse component from LR+SD. For each 
dataset is shown (top row) TFR over a frequency range about the alpha band and (bottom row) the signal strength at 10Hz with single standard deviation bounds.}
\label{fig:tfig}
\end{figure}

Figure \ref{fig:tfig} shows single-patient alpha power averaged over all stimuli for a window of 2sec pre stimulus to 8sec post. The raw data is shown in comparison with the sparse component 
from LR+SD and ICA using an average of the time-frequency intensity over alpha band frequencies. The strength at the specific frequency of 10Hz with bounds of one standard deviation is also 
shown for each dataset.

\section{Discussion}

In the present paper, we introduce a new method for removing artifacts from EEG signals, which decomposes data into the sum of two matrices: a sparse matrix representing the cleaned data, and 
a low-rank matrix, which corresponds to the artifact portion of the data.  We applied this algorithm to remove artifact from two forms of data. In the first, we used simulated data where the 
pure EEG data was known, and corrupted by the addition of a series of canonical artifacts to the data using a randomized mixing matrix.  In the second set of data, we applied LR+SD to 
empirical EEG data collected concurrently with MRI data.  

Overall, LR+SD appeared effective on each of the two forms of event related data. The LR+SD algorithm was very effective at separating artifact from the pure EEG signal in our simulated data 
set with up to three true EEG sources. Our method also recovered the diminution in alpha power following a subject viewing a Gabor light flash, and improved the signal to noise ratio of 
detecting this ERSP by ~34\%. The LR+SD signal was clearly not as robust as the data collected out of scanner. Visually, the diminution in alpha power became apparent upon inspection 
following LR+SD in topographic projections of the data as well as at the singe channel sensor level on the scalp. 

\section{Examining BCG Removal \& Sparsity}

Topographically, the BCG artifact is both non-uniform and dynamic. Its complex signature is thought to derive primarily from three key origins each with distinct R-wave latencies: 
pulse-driven head rotation, the Hall effect due to pulsatile blood flow, and pulsatile expansion of the scalp \citep{Mullinger2013}. If the BCG artifact was spatially static, it is possible 
that channel by channel template based methods may be adequate for its removal. However, the non stationary nature of the BCG makes its characterization and removal more complicated. 
Categorically, the pulse-driven head rotation appears to contribute most prominently to the amplitude of the artifact signal \citep{Mullinger2013}; however, the artifact, despite its origin, 
appears to be reflected in nearly all channels to varying degrees. 

Overall, the LR+SD algorithm is quite similar to the robust PCA algorithm, which relies on a number of assumptions about the data. Specifically, it is assumed that the data events themselves 
are sparsely represented in the time domain. Artifacts are conversely assumed to be broadly distributed at the channel level across the scalp, which appears to be a relevant assumption given 
what has been observed about the distribution of EEG artifacts, mentioned above, thus far. 

Results using ICA for artifact removal of concurrent EEG-fMRI have varied \citep{Mantini2007,Eichele2005}. Interestingly, the BCG artifact and its topographic variability appear to be more 
problematic when experiments are carried out at higher MR field strengths. Results using ICA as a method for BCG artifact removal at 1.5T were initially quite promising 
\citep{Mantini2007,Eichele2005}. However, ICA has been shown to fail at 3T \citep{Debener2008}, the field strength used in the present study. It is possible that the larger magnitude artifact 
at higher field strength amplifies subtleties of the artifact that are subsequently split across components. Compared to ICA, LR+SD cleaning appeared to significantly improved the SNR of 
detectable ERSPs by controlling the rank of the artifact matrix.  

In each of these data sets, the events occurred in temporally distinct time points relative to the data segment or epoch, making the sparsity assumption in the temporal domain relevant. 
Further, the ERSPs for the experimental data were measured across a dense 256 electrode net, and voltage fluctuations due to flash stimuli in channels located at large spatial distances from 
occipital electrodes is likely quite minimal. In this sense, the data of interest were also represented across a sparse number of channels. Further work is needed to characterize the 
conditions under which this sparsity assumption holds.  For example, it is unclear how well this algorithm would perform when the signal of interest is continuously present across distributed 
electrodes, as might be the case when examining default mode network activity \citep{Biswal1995}.

It is also interesting to note that, after optimization, we found that setting our regularization parameter to rank 25 was effective at isolating BCG artifact. Lower ranks were ineffective at 
isolating the BCG noise, due to the complexity of the BCG signal itself. In the clinical setting, the ECG signal is often low-pass filtered and gross changes of its signature (e.g. ST segment 
elevation) are examined.  However, lower amplitude changes in higher frequencies of the ECG exist and have been shown to correlate with cardiac ischemia \citep{Schlegel2004} even up to 250Hz 
\citep{Douglas2006}. Given this broad spectral signature, it is unsurprising that more that a few sparse components were required to capture the artifact signals. In our online supplemental 
results, we provide a video “sweep,” illustrating how the low rank matrix, thought to reflect BCG and other artifacts,  and sparse matrix, thought to reflect the data, components vary 
according to our regularization parameter value (see Supplemental).

\section{Conclusions and Future Work }

In this paper, we demonstrate that LR+SD worked quite well on event related data that was either simulated or empirical. One of the strengths of this method is that it eliminates the need for 
burdensome manual examination of components, and expert used decisions required to determine whether a given component is noise or part of the data themselves. This process, as may be the 
case when using ICA based artifact removal, can also vary across users. LR+SD provides an automatic method for parsing data and artifact into separate groups, with the need to tune only one 
regularization parameter. Nonetheless, the SNR of data post LR+SD was still less that that of data collected outside the scanning environment. Future work may include applying this algorithm 
in combination with temporal kernel methods (e.g. \cite{Biessmann2010}) to effectively map the dynamic artifact to a stationary reference frame, and consider additional types of event related 
neuroimaging data that are corrupted by distributed artifact effects. 

\section{Acknowledgements}
The authors want to thank the Keck Foundation for their support.

\bibliographystyle{model2-names}
\bibliography{references_PKD}

\begin{thebibliography}{36}
\expandafter\ifx\csname natexlab\endcsname\relax\def\natexlab#1{#1}\fi
\providecommand{\url}[1]{\texttt{#1}}
\providecommand{\href}[2]{#2}
\providecommand{\path}[1]{#1}
\providecommand{\DOIprefix}{doi:}
\providecommand{\ArXivprefix}{arXiv:}
\providecommand{\URLprefix}{URL: }
\providecommand{\Pubmedprefix}{pmid:}
\providecommand{\doi}[1]{\href{http://dx.doi.org/#1}{\path{#1}}}
\providecommand{\Pubmed}[1]{\href{pmid:#1}{\path{#1}}}
\providecommand{\bibinfo}[2]{#2}
\ifx\xfnm\relax \def\xfnm[#1]{\unskip,\space#1}\fi
\bibitem[{Allen et~al.(2000)Allen, Josephs and Turner}]{Allen2000}
\bibinfo{author}{Allen, P.J.}, \bibinfo{author}{Josephs, O.}, \bibinfo{author}{Turner, R.}, \bibinfo{year}{2000}.
\newblock \bibinfo{title}{A method for removing imaging artifact from continuous eeg recorded during functional mri}.
\newblock \bibinfo{journal}{NeuroImage} \bibinfo{volume}{12}, \bibinfo{pages}{230--239}.
\bibitem[{Allen et~al.(1998)Allen, Polizzi, Krakow, Fish and Lemieux}]{Allen1998}
\bibinfo{author}{Allen, P.J.}, \bibinfo{author}{Polizzi, G.}, \bibinfo{author}{Krakow, K.}, \bibinfo{author}{Fish, D.R.}, \bibinfo{author}{Lemieux, L.}, \bibinfo{year}{1998}.
\newblock \bibinfo{title}{Identification of {EEG} events in the {MR} scanner: the problem of pulse artifact and a method for its subtraction}.
\newblock \bibinfo{journal}{NeuroImage} \bibinfo{volume}{8}, \bibinfo{pages}{229--239}.
\bibitem[{Biessmann et~al.(2010)Biessmann, Meinecke, Gretton, Rauch, Rainer, Logothetis and Muller}]{Biessmann2010}
\bibinfo{author}{Biessmann, F.}, \bibinfo{author}{Meinecke, C.}, \bibinfo{author}{Gretton, A.}, \bibinfo{author}{Rauch, A.}, \bibinfo{author}{Rainer, G.}, \bibinfo{author}{Logothetis, N.}, \bibinfo{author}{Muller, K.}, \bibinfo{year}{2010}.
\newblock \bibinfo{title}{Temporal kernel cca and its application in multimodal neuronal data analysis}.
\newblock \bibinfo{journal}{Machine Learning} \bibinfo{volume}{79}, \bibinfo{pages}{5--27}.
\bibitem[{Biswal et~al.(1995)Biswal, Yetkin, Haughton and Hyde}]{Biswal1995}
\bibinfo{author}{Biswal, B.}, \bibinfo{author}{Yetkin, F.}, \bibinfo{author}{Haughton, V.}, \bibinfo{author}{Hyde, J.}, \bibinfo{year}{1995}.
\newblock \bibinfo{title}{Functional connectivity in the motor cortex of resting human brain using echo-planar mri}.
\newblock \bibinfo{journal}{Magn Reson Med} \bibinfo{volume}{34(4)}, \bibinfo{pages}{537--541}.
\bibitem[{Buzsáki et~al.(2012)Buzsáki, Anastassiou and Koch}]{Buzsaki2012}
\bibinfo{author}{Buzsáki, G.}, \bibinfo{author}{Anastassiou, C.}, \bibinfo{author}{Koch, C.}, \bibinfo{year}{2012}.
\newblock \bibinfo{title}{The origin of extracellular fields and currents — eeg, ecog, lfp and spikes}.
\newblock \bibinfo{journal}{Nature Reviews Neuroscience} \bibinfo{volume}{13(6)}, \bibinfo{pages}{407--420}.
\bibitem[{Cai et~al.(2008)Cai, Cand\`es and Shen}]{Cai2008}
\bibinfo{author}{Cai, J.F.}, \bibinfo{author}{Cand\`es, E.}, \bibinfo{author}{Shen, Z.}, \bibinfo{year}{2008}.
\newblock \bibinfo{title}{A singular value thresholding algorithm for matrix completion}.
\newblock \bibinfo{journal}{SIAM Journal on Optimization} \bibinfo{volume}{20}, \bibinfo{pages}{1956--1982}.
\bibitem[{Cand\`es et~al.(2009)Cand\`es, Li, Ma and Wright}]{Candes2009}
\bibinfo{author}{Cand\`es, E.}, \bibinfo{author}{Li, X.}, \bibinfo{author}{Ma, Y.}, \bibinfo{author}{Wright, J.}, \bibinfo{year}{2009}.
\newblock \bibinfo{title}{Robust principal component analysis}.
\newblock \bibinfo{journal}{Journal of the ACM} \bibinfo{volume}{58}, \bibinfo{pages}{1--37}.
\bibitem[{Correa et~al.(2007)Correa, Laciar, Pati$\tilde{\text{n}}$o and Valentinuzzi}]{Correa2007}
\bibinfo{author}{Correa, A.G.}, \bibinfo{author}{Laciar, E.}, \bibinfo{author}{Pati$\tilde{\text{n}}$o, H.D.}, \bibinfo{author}{Valentinuzzi, M.E.}, \bibinfo{year}{2007}.
\newblock \bibinfo{title}{Artifact removal from {EEG} signals using adaptive filters in cascade}, in: \bibinfo{booktitle}{16th Argentine Bioengineering Congress and the 5th Conference of Clinical Engineering}.
\bibitem[{Dahne et~al.(2013)Dahne, Biessman, Meinecke, Mehnert, Fazli and Muller}]{Dahne2012}
\bibinfo{author}{Dahne, S.}, \bibinfo{author}{Biessman, F.}, \bibinfo{author}{Meinecke, F.}, \bibinfo{author}{Mehnert, J.}, \bibinfo{author}{Fazli, S.}, \bibinfo{author}{Muller, K.}, \bibinfo{year}{2013}.
\newblock \bibinfo{title}{Integration of multivariate data streams with bandpower signals}.
\newblock \bibinfo{journal}{IEEE Transactions on Multimedia} \bibinfo{volume}{15(5)}, \bibinfo{pages}{1001--1013}.
\bibitem[{Daunizeau et~al.(2009)Daunizeau, Laufs and Friston}]{Daunizeau2009}
\bibinfo{author}{Daunizeau, J.}, \bibinfo{author}{Laufs, H.}, \bibinfo{author}{Friston, K.}, \bibinfo{year}{2009}.
\newblock \bibinfo{title}{Eeg–fmri information fusion: Biophysics and data analysis}, in: \bibinfo{editor}{Mulert, C.}, \bibinfo{editor}{Lemieux, L.} (Eds.), \bibinfo{booktitle}{EEG-fMRI}. \bibinfo{publisher}{{Springer} {P}ress}. volume~\bibinfo{volume}{1}, pp. \bibinfo{pages}{511--526}.
\bibitem[{Debener et~al.(2008)Debener, Mullinger, Niazy and Bowtell}]{Debener2008}
\bibinfo{author}{Debener, S.}, \bibinfo{author}{Mullinger, K.}, \bibinfo{author}{Niazy, R.}, \bibinfo{author}{Bowtell, R.}, \bibinfo{year}{2008}.
\newblock \bibinfo{title}{Properties of the ballistocardiogram artifact as revealed by eeg recordings at 1.5, 3, and 7t static magnetic field strength}.
\newblock \bibinfo{journal}{Int J Psychophysiol} \bibinfo{volume}{67(3)}, \bibinfo{pages}{189--99}.
\bibitem[{Debener et~al.(2007)Debener, Strobel, Sorger, Peters, Kranczioch, Engel and Goebel}]{Debener2007}
\bibinfo{author}{Debener, S.}, \bibinfo{author}{Strobel, A.}, \bibinfo{author}{Sorger, B.}, \bibinfo{author}{Peters, J.}, \bibinfo{author}{Kranczioch, C.}, \bibinfo{author}{Engel, A.}, \bibinfo{author}{Goebel, R.}, \bibinfo{year}{2007}.
\newblock \bibinfo{title}{Improved quality of auditory event-related potentials recorded simultaneously with 3-t fmri: removal of the ballistocardiogram artefact.}
\newblock \bibinfo{journal}{NeuroImage} \bibinfo{volume}{34(2)}, \bibinfo{pages}{587--97}.
\bibitem[{Douglas et~al.(2006)Douglas, Batdorf, Evans, Feiveson, Arenare and Schlegel}]{Douglas2006}
\bibinfo{author}{Douglas, P.}, \bibinfo{author}{Batdorf, N.}, \bibinfo{author}{Evans, R.}, \bibinfo{author}{Feiveson, A.}, \bibinfo{author}{Arenare, B.}, \bibinfo{author}{Schlegel, T.}, \bibinfo{year}{2006}.
\newblock \bibinfo{title}{Temporal and postural variation of 12-lead high-frequency qrs electrocardiographic signals in asymptomatic individuals}.
\newblock \bibinfo{journal}{J Electrocardiol} \bibinfo{volume}{39(3)}, \bibinfo{pages}{259--265}.
\bibitem[{Eichele et~al.(2005)Eichele, Specht, Moosmann, Jongsma, Quiroga, Nordby and Hugdahl}]{Eichele2005}
\bibinfo{author}{Eichele, T.}, \bibinfo{author}{Specht, K.}, \bibinfo{author}{Moosmann, M.}, \bibinfo{author}{Jongsma, M.}, \bibinfo{author}{Quiroga, R.}, \bibinfo{author}{Nordby, H.}, \bibinfo{author}{Hugdahl, K.}, \bibinfo{year}{2005}.
\newblock \bibinfo{title}{Assessing the spatiotemporal evolution of neuronal activation with single-trial event-related potentials and functional mri.}
\newblock \bibinfo{journal}{PNAS} \bibinfo{volume}{102(49)}, \bibinfo{pages}{17798--803}.
\bibitem[{Goldberger et~al.(2000)Goldberger, Amaral, Glass, Hausdorff, Ivanov, Mark, Mietus, Moody, Peng and Stanley}]{Physionet}
\bibinfo{author}{Goldberger, A.L.}, \bibinfo{author}{Amaral, L.A.N.}, \bibinfo{author}{Glass, L.}, \bibinfo{author}{Hausdorff, J.M.}, \bibinfo{author}{Ivanov, P.C.}, \bibinfo{author}{Mark, R.G.}, \bibinfo{author}{Mietus, J.E.}, \bibinfo{author}{Moody, G.B.}, \bibinfo{author}{Peng, C.K.}, \bibinfo{author}{Stanley, H.}, \bibinfo{year}{2000}.
\newblock \bibinfo{title}{{PhysioBank, PhysioToolkit, and PhysioNet}: Components of a new research resource for complex physiologic signals}.
\newblock \bibinfo{journal}{Circulation} \bibinfo{volume}{101}, \bibinfo{pages}{e215--e220}.
\bibitem[{Goldman et~al.(2002)Goldman, Stern, Engel and Cohen}]{Goldman2002}
\bibinfo{author}{Goldman, R.}, \bibinfo{author}{Stern, J.}, \bibinfo{author}{Engel, J.}, \bibinfo{author}{Cohen, M.}, \bibinfo{year}{2002}.
\newblock \bibinfo{title}{Simultaneous eeg and fmri of the alpha rhythm}.
\newblock \bibinfo{journal}{Neuroreport} \bibinfo{volume}{13(18)}, \bibinfo{pages}{2487--2492}.
\bibitem[{Grouiller et~al.(2011)Grouiller, Thornton, Groening, Spinelli, Duncan, Schaller, Siniatchin, Lemieux, Seeck, Michel and Vulliemoz}]{Grouiller2011}
\bibinfo{author}{Grouiller, F.}, \bibinfo{author}{Thornton, R.}, \bibinfo{author}{Groening, K.}, \bibinfo{author}{Spinelli, L.}, \bibinfo{author}{Duncan, J.}, \bibinfo{author}{Schaller, K.}, \bibinfo{author}{Siniatchin, M.}, \bibinfo{author}{Lemieux, L.}, \bibinfo{author}{Seeck, M.}, \bibinfo{author}{Michel, C.M.}, \bibinfo{author}{Vulliemoz, S.}, \bibinfo{year}{2011}.
\newblock \bibinfo{title}{With or without spikes: localization of focal epileptic activity by simultaneous electroencephalography and functional magnetic resonance imaging}.
\newblock \bibinfo{journal}{Brain} \bibinfo{volume}{134(10)}, \bibinfo{pages}{2867--2886}.
\bibitem[{Huettel et~al.(2004)Huettel, McKeown, Song, Hart, Spencer, Allison and McCarthy}]{Huettel2004}
\bibinfo{author}{Huettel, S.}, \bibinfo{author}{McKeown, M.}, \bibinfo{author}{Song, A.}, \bibinfo{author}{Hart, S.}, \bibinfo{author}{Spencer, D.}, \bibinfo{author}{Allison, T.}, \bibinfo{author}{McCarthy, G.}, \bibinfo{year}{2004}.
\newblock \bibinfo{title}{Linking hemodynamic and electrophysiological measures of brain activity: evidence from functional mri and intracranial field potentials}.
\newblock \bibinfo{journal}{Cerebral Cortex} \bibinfo{volume}{14(2)}, \bibinfo{pages}{165--173}.
\bibitem[{Jung et~al.(1998)Jung, Humphries, Lee, Makeig, McKeown, Iragui and Sejnowski}]{Jung1998}
\bibinfo{author}{Jung, T.P.}, \bibinfo{author}{Humphries, C.}, \bibinfo{author}{Lee, T.W.}, \bibinfo{author}{Makeig, S.}, \bibinfo{author}{McKeown, M.J.}, \bibinfo{author}{Iragui, V.}, \bibinfo{author}{Sejnowski, T.J.}, \bibinfo{year}{1998}.
\newblock \bibinfo{title}{Removing electroencephalographic artifacts : Comparison between {ICA} and {PCA}}.
\newblock \bibinfo{journal}{Neural Networks for Signal Processing} \bibinfo{volume}{VIII}, \bibinfo{pages}{63--72}.
\bibitem[{Jung et~al.(2000)Jung, Makeig, Humphries, Lee, McKeown, Iragui and Sejnowski}]{Jung2000}
\bibinfo{author}{Jung, T.P.}, \bibinfo{author}{Makeig, S.}, \bibinfo{author}{Humphries, C.}, \bibinfo{author}{Lee, T.W.}, \bibinfo{author}{McKeown, M.J.}, \bibinfo{author}{Iragui, V.}, \bibinfo{author}{Sejnowski, T.J.}, \bibinfo{year}{2000}.
\newblock \bibinfo{title}{Removing electroencephalographic artifacts by blind source separation}.
\newblock \bibinfo{journal}{Psychophysiology} \bibinfo{volume}{37}, \bibinfo{pages}{163--178}.
\bibitem[{Lin et~al.(2009)Lin, Chen, Wu and Ma}]{Lin2009}
\bibinfo{author}{Lin, Z.}, \bibinfo{author}{Chen, M.}, \bibinfo{author}{Wu, L.}, \bibinfo{author}{Ma, Y.}, \bibinfo{year}{2009}.
\newblock \bibinfo{title}{The Augmented Lagrange Multiplier Method for Exact Recovery of Corrupted {L}ow-{R}ank Matrices}.
\newblock \bibinfo{type}{Technical Report}. University of Illinois at Urbana-Champaign.
\newblock \bibinfo{note}{UILU-ENG-09-2215, arXiv:1009.5055v2}.
\bibitem[{Lind et~al.(2003)Lind, Goodwin, Hill, Ali, Redline and Quan}]{SHHS2}
\bibinfo{author}{Lind, B.}, \bibinfo{author}{Goodwin, J.}, \bibinfo{author}{Hill, J.}, \bibinfo{author}{Ali, T.}, \bibinfo{author}{Redline, S.}, \bibinfo{author}{Quan, S.}, \bibinfo{year}{2003}.
\newblock \bibinfo{title}{{Recruitment of healthy adults into a study of overnight sleep monitoring in the home: experience of The Sleep Heart Health Study}}.
\newblock \bibinfo{journal}{Sleep Breath} \bibinfo{volume}{7}, \bibinfo{pages}{13--24}.
\bibitem[{Makeig et~al.(1996)Makeig, Bell, Jung and Sejnowski}]{Makeig1996}
\bibinfo{author}{Makeig, S.}, \bibinfo{author}{Bell, A.J.}, \bibinfo{author}{Jung, T.P.}, \bibinfo{author}{Sejnowski, T.J.}, \bibinfo{year}{1996}.
\newblock \bibinfo{title}{Independent {C}omponent {A}nalysis of {E}lectroencephalographic data}, in: \bibinfo{editor}{Touretzky, D.}, \bibinfo{editor}{Mozer, M.}, \bibinfo{editor}{Hasselmo, M.} (Eds.), \bibinfo{booktitle}{Advances in Neural Information Processing Systems}. \bibinfo{publisher}{{MIT} {P}ress}. volume~\bibinfo{volume}{8}, pp. \bibinfo{pages}{145--151}.
\bibitem[{Mantini et~al.(2007)Mantini, Perrucci, Cugini, Ferretti, Romani and Gratta}]{Mantini2007}
\bibinfo{author}{Mantini, D.}, \bibinfo{author}{Perrucci, M.}, \bibinfo{author}{Cugini, S.}, \bibinfo{author}{Ferretti, A.}, \bibinfo{author}{Romani, G.}, \bibinfo{author}{Gratta, C.D.}, \bibinfo{year}{2007}.
\newblock \bibinfo{title}{Complete artifact removal for eeg recorded during continuous {fMRI} using independent component analysis}.
\newblock \bibinfo{journal}{NeuroImage} \bibinfo{volume}{34}, \bibinfo{pages}{598--607}.
\bibitem[{Masterton et~al.(2007)Masterton, Abbott, Fleming and Jackson}]{Masterton2007}
\bibinfo{author}{Masterton, R.A.}, \bibinfo{author}{Abbott, D.F.}, \bibinfo{author}{Fleming, S.W.}, \bibinfo{author}{Jackson, G.D.}, \bibinfo{year}{2007}.
\newblock \bibinfo{title}{Measurement and reduction of motion and ballistocardiogram artefacts from simultaneous {EEG} and {fMRI} recordings}.
\newblock \bibinfo{journal}{NeuroImage} \bibinfo{volume}{37}, \bibinfo{pages}{202--211}.
\bibitem[{Mullinger et~al.(2013)Mullinger, Havenhand and Bowtell}]{Mullinger2013}
\bibinfo{author}{Mullinger, K.}, \bibinfo{author}{Havenhand, J.}, \bibinfo{author}{Bowtell, R.}, \bibinfo{year}{2013}.
\newblock \bibinfo{title}{Identifying the sources of the pulse artifact in eeg recordings made inside the mri scanner}.
\newblock \bibinfo{journal}{NeuroImage} \bibinfo{volume}{71(1)}, \bibinfo{pages}{75--83}.
\bibitem[{Niazy et~al.(2005)Niazy, Beckmann, Iannetti, Brady and Smith}]{Niazy2005}
\bibinfo{author}{Niazy, R.}, \bibinfo{author}{Beckmann, C.}, \bibinfo{author}{Iannetti, G.}, \bibinfo{author}{Brady, J.}, \bibinfo{author}{Smith, S.}, \bibinfo{year}{2005}.
\newblock \bibinfo{title}{Removal of fmri environment artifacts from eeg data using optimal basis sets}.
\newblock \bibinfo{journal}{NeuroImage} \bibinfo{volume}{28(3)}, \bibinfo{pages}{720--737}.
\bibitem[{Quan et~al.(1997)Quan, Howard, Iber, Kiley, Nieto, O'Connor, Rapoport, Redline, Robbins, JM and Wahl}]{SHHS1}
\bibinfo{author}{Quan, S.}, \bibinfo{author}{Howard, B.}, \bibinfo{author}{Iber, C.}, \bibinfo{author}{Kiley, J.}, \bibinfo{author}{Nieto, F.}, \bibinfo{author}{O'Connor, G.}, \bibinfo{author}{Rapoport, D.}, \bibinfo{author}{Redline, S.}, \bibinfo{author}{Robbins, J.}, \bibinfo{author}{JM, S.}, \bibinfo{author}{Wahl, P.}, \bibinfo{year}{1997}.
\newblock \bibinfo{title}{{The Sleep Heart Health Study: design, rationale, and methods}}.
\newblock \bibinfo{journal}{Sleep} \bibinfo{volume}{20}, \bibinfo{pages}{1077--1085}.
\bibitem[{Redline et~al.(1998)Redline, Sander, Lind, Quan, Iber, Gottlieb, Bonekat, Rapoport, Smith and Kiley}]{SHHS3}
\bibinfo{author}{Redline, S.}, \bibinfo{author}{Sander, M.}, \bibinfo{author}{Lind, B.}, \bibinfo{author}{Quan, S.}, \bibinfo{author}{Iber, C.}, \bibinfo{author}{Gottlieb, D.}, \bibinfo{author}{Bonekat, W.}, \bibinfo{author}{Rapoport, D.}, \bibinfo{author}{Smith, P.}, \bibinfo{author}{Kiley, J.}, \bibinfo{year}{1998}.
\newblock \bibinfo{title}{{Methods for obtaining and analyzing unattended polysomnography data for a multicenter study}}.
\newblock \bibinfo{journal}{Sleep} \bibinfo{volume}{21}, \bibinfo{pages}{759--767}.
\bibitem[{Ritter and Villringer(2006)}]{Ritter2006}
\bibinfo{author}{Ritter, P.}, \bibinfo{author}{Villringer, A.}, \bibinfo{year}{2006}.
\newblock \bibinfo{title}{Simultaneous eeg-fmri}.
\newblock \bibinfo{journal}{Neuroscience and biobehavioral reviews} \bibinfo{volume}{30(6)}, \bibinfo{pages}{823--838}.
\bibitem[{R.Thornton et~al.(2011)R.Thornton, Vulliemoz, Rodionov, Carmichael, Chaudhary, Diehl, Laufs, Vollmar, McEvoy, Walker, Bartolomei, Guye, Chauvel, Duncan and Lemieux.}]{Thornton2011}
\bibinfo{author}{R.Thornton}, \bibinfo{author}{Vulliemoz, S.}, \bibinfo{author}{Rodionov, R.}, \bibinfo{author}{Carmichael, D.}, \bibinfo{author}{Chaudhary, U.}, \bibinfo{author}{Diehl, B.}, \bibinfo{author}{Laufs, H.}, \bibinfo{author}{Vollmar, C.}, \bibinfo{author}{McEvoy, A.}, \bibinfo{author}{Walker, M.}, \bibinfo{author}{Bartolomei, F.}, \bibinfo{author}{Guye, M.}, \bibinfo{author}{Chauvel, P.}, \bibinfo{author}{Duncan, J.}, \bibinfo{author}{Lemieux., L.}, \bibinfo{year}{2011}.
\newblock \bibinfo{title}{Epileptic networks in focal cortical dysplasia revealed using electroencephalography-functional magnetic resonance imaging.}
\newblock \bibinfo{journal}{NeuroImage} \bibinfo{volume}{70(5)}, \bibinfo{pages}{822--837}.
\bibitem[{Schlegel et~al.(2004)Schlegel, Kulecz, DePalma, Feiveson, Wilson, Rahman and Bungo}]{Schlegel2004}
\bibinfo{author}{Schlegel, T.}, \bibinfo{author}{Kulecz, W.}, \bibinfo{author}{DePalma, J.}, \bibinfo{author}{Feiveson, A.}, \bibinfo{author}{Wilson, J.}, \bibinfo{author}{Rahman, M.}, \bibinfo{author}{Bungo, M.}, \bibinfo{year}{2004}.
\newblock \bibinfo{title}{Real-time 12-lead high-frequency qrs electrocardiography for enhanced detection of myocardial ischemia and coronary artery disease}.
\newblock \bibinfo{journal}{Mayo Clinic Proceedings} \bibinfo{volume}{79(3)}, \bibinfo{pages}{339--350}.
\bibitem[{Srivastava et~al.(2005)Srivastava, Crottaz-Herbette, Lau, Glover and Menon}]{Srivastava2005}
\bibinfo{author}{Srivastava, G.}, \bibinfo{author}{Crottaz-Herbette, S.}, \bibinfo{author}{Lau, K.}, \bibinfo{author}{Glover, G.}, \bibinfo{author}{Menon, V.}, \bibinfo{year}{2005}.
\newblock \bibinfo{title}{Ica-based procedures for removing ballistogardiogram artifacts from eeg data acquired in the mri scanner}.
\newblock \bibinfo{journal}{Neuroimage} \bibinfo{volume}{24(1)}, \bibinfo{pages}{50--60}.
\bibitem[{Sun et~al.(2009)Sun, Rieger and Hinrichs}]{Sun2009}
\bibinfo{author}{Sun, L.}, \bibinfo{author}{Rieger, J.}, \bibinfo{author}{Hinrichs, H.}, \bibinfo{year}{2009}.
\newblock \bibinfo{title}{Maximum noise fraction ({MNF}) transformation to remove ballistocardiographic artifacts in {EEG} signals recorded during {fMRI} scanning}.
\newblock \bibinfo{journal}{NeuroImage} \bibinfo{volume}{46}, \bibinfo{pages}{144--153}.
\bibitem[{Valdes-Sosa et~al.(2009)Valdes-Sosa, Sanchez-Bornot, Sotero, Iturria-Medina, Aleman-Gomez, Bosch-Bayard and Ozaki}]{ValdesSosa2009}
\bibinfo{author}{Valdes-Sosa, P.}, \bibinfo{author}{Sanchez-Bornot, J.}, \bibinfo{author}{Sotero, R.}, \bibinfo{author}{Iturria-Medina, Y.}, \bibinfo{author}{Aleman-Gomez, Y.}, \bibinfo{author}{Bosch-Bayard, J.}, \bibinfo{author}{Ozaki, T.}, \bibinfo{year}{2009}.
\newblock \bibinfo{title}{Model driven eeg/fmri fusion of brain oscillations}.
\newblock \bibinfo{journal}{Human Brain Mapping} \bibinfo{volume}{30(9)}, \bibinfo{pages}{2701--2721}.
\bibitem[{Zou et~al.(2012)Zou, Jr. and Jafari}]{Zou2012}
\bibinfo{author}{Zou, Y.}, \bibinfo{author}{Jr., J.H.}, \bibinfo{author}{Jafari, R.}, \bibinfo{year}{2012}.
\newblock \bibinfo{title}{Automatic {EEG} artifact removal based on {ICA} and hierarchical clustering}, in: \bibinfo{booktitle}{2012 IEEE International Conference on Acoustics, Speech and Signal Processing (ICASSP)}, \bibinfo{address}{Kyoto}. pp. \bibinfo{pages}{649--652}.

\end{thebibliography}

\appendix
\section{Additional Simulation Data}

\begin{figure}[!h]
\includegraphics[width=\columnwidth,height=0.7\columnwidth]{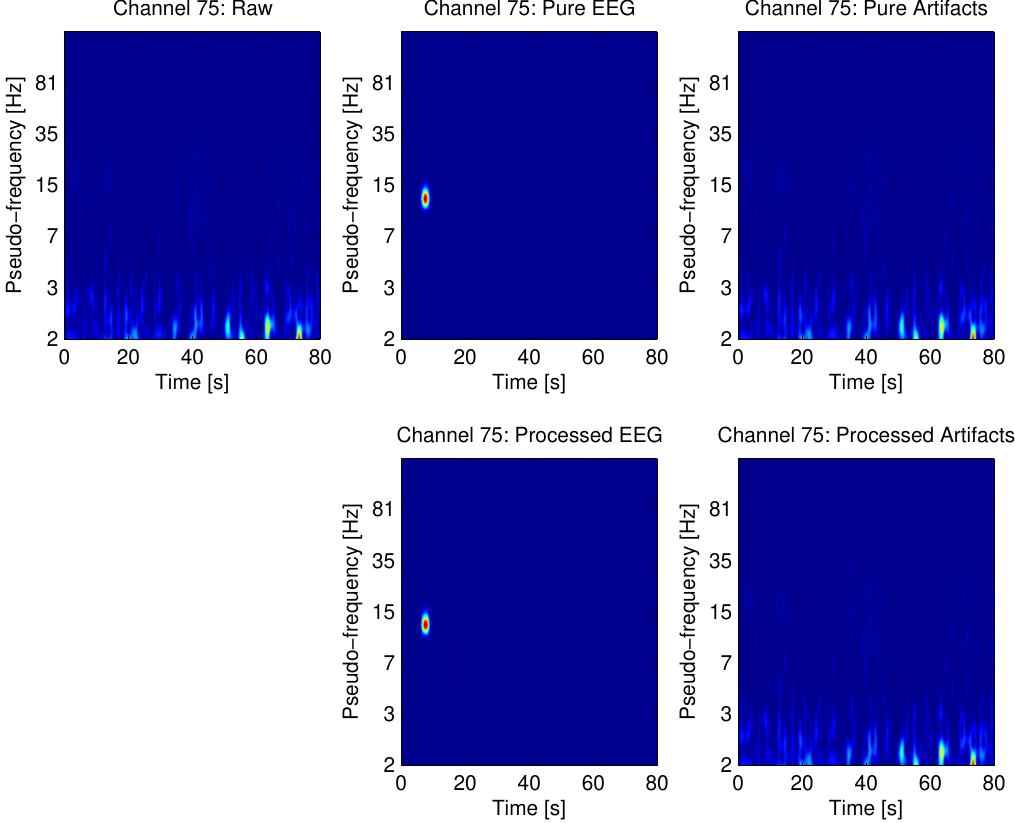}
\caption{
Simulated Data Set II, second electrode. Three sources measured with an electrode located close to the visual cortex. (Top Panel) CWT of original simulated data consisting of three dipole sources corrupted 
by ECG, EMG and right and left EOG template artifacts (left), the true EEG data alone (middle), and artifact alone (right). (Bottom Panel) LR+SD result TFRs separating the cleaned EEG (left) 
from artifact (right). Values are normalized to maximum for each display.
}
\label{fig:three75}
\end{figure}

\begin{figure}[!h]
\includegraphics[width=\columnwidth,height=0.7\columnwidth]{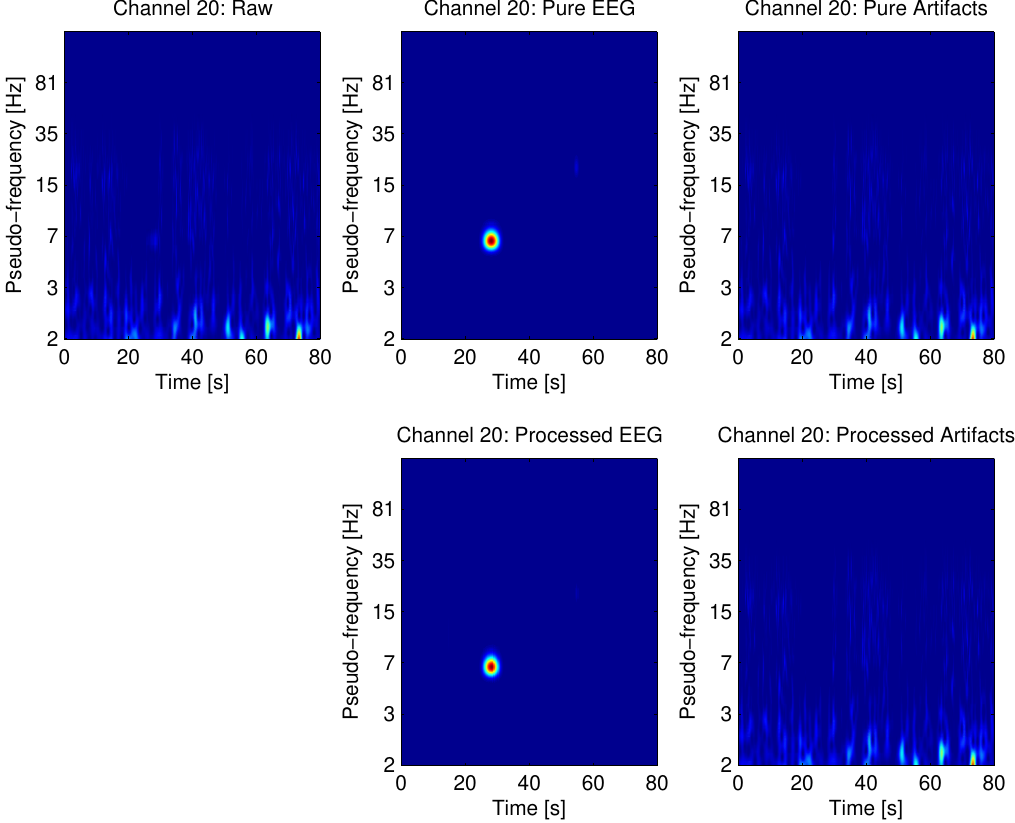}
\caption{
Simulated Data Set II, third electrode. Three sources measured with an electrode located close to the frontal lobe area. (Top Panel) CWT of original simulated data consisting of three dipole 
sources corrupted by ECG, EMG and right and left EOG template artifacts (left), the true EEG data alone (middle), and artifact alone (right). (Bottom Panel) LR+SD result TFRs separating the 
cleaned EEG (left) from artifact (right). Values are normalized to maximum for each display.
}
\label{fig:three20}
\end{figure}

We simulated an EEG signal using three sources mixed with multiple artifacts. The analysis of the resultant data can be performed on any particular electrode. In Figures \ref{fig:three75}
and \ref{fig:three20} we demonstrate the same analysis as in Figure \ref{fig:three37} but instead using, respectively, data from the simulated electrodes near the visual cortex and close to the frontal 
lobe area.

\end{document}